\documentclass[twocolumn,prl,aps,superscriptaddress]{revtex4-1}
\usepackage[latin9]{inputenc}
\usepackage{mathtools}
\usepackage{amsbsy}
\usepackage{amstext}
\usepackage[unicode=true,pdfusetitle,
 bookmarks=false,
 breaklinks=false,pdfborder={0 0 1},backref=false,colorlinks=false]
 {hyperref}
\hypersetup{
 bookmarksnumbered=false,bookmarksopen=false}
\makeatletter

\@ifundefined{textcolor}{}
{%
 \definecolor{BLACK}{gray}{0}
 \definecolor{WHITE}{gray}{1}
 \definecolor{RED}{rgb}{1,0,0}
 \definecolor{GREEN}{rgb}{0,1,0}
 \definecolor{BLUE}{rgb}{0,0,1}
 \definecolor{CYAN}{cmyk}{1,0,0,0}
 \definecolor{MAGENTA}{cmyk}{0,1,0,0}
 \definecolor{YELLOW}{cmyk}{0,0,1,0}
}

\usepackage{amsmath}
\usepackage{graphicx}
\usepackage{amssymb}
\usepackage{txfonts,color}
\makeatother

\begin{document}
\title{Selective Interactions in the Quantum Rabi Model}
\author{L. Cong}
\affiliation{International Center of Quantum Artificial Intelligence for Science and Technology~(QuArtist)
and Department of Physics, Shanghai University, 200444 Shanghai, China}
\author{S. Felicetti}
\affiliation{Departamento de F\'isica Te\'orica de la Materia Condensada and Condensed Matter
Physics Center (IFIMAC), Universidad Aut\'onoma de Madrid, 28049 Madrid, Spain}
\author{J. Casanova}
\affiliation{Department of Physical Chemistry, University of the Basque Country UPV/EHU, Apartado 644, 48080 Bilbao, Spain}
\affiliation{IKERBASQUE,  Basque  Foundation  for  Science,  Maria  Diaz  de  Haro  3,  48013  Bilbao,  Spain}
\author{L. Lamata}
\affiliation{Department of Physical Chemistry, University of the Basque Country UPV/EHU, Apartado 644, 48080 Bilbao, Spain}
\affiliation{Departamento de F\'isica At\'omica, Molecular y Nuclear, Universidad de Sevilla, 41080 Sevilla, Spain}
\author{E. Solano}
\affiliation{International Center of Quantum Artificial Intelligence for Science and Technology~(QuArtist)
and Department of Physics, Shanghai University, 200444 Shanghai, China}
\affiliation{Department of Physical Chemistry, University of the Basque Country UPV/EHU, Apartado 644, 48080 Bilbao, Spain}
\affiliation{IKERBASQUE,  Basque  Foundation  for  Science,  Maria  Diaz  de  Haro  3,  48013  Bilbao,  Spain}
\author{I. Arrazola}
\affiliation{Department of Physical Chemistry, University of the Basque Country UPV/EHU, Apartado 644, 48080 Bilbao, Spain}

\begin{abstract}
We demonstrate the emergence of selective $k$-photon interactions in the strong and ultrastrong coupling regimes of the quantum Rabi model with a Stark coupling term. In particular, we show that the interplay between the rotating and counter-rotating terms produces multi-photon interactions whose resonance frequencies depend, due to the Stark term, on the state of the bosonic mode. We develop an analytical framework to explain these $k$-photon interactions by using time-dependent perturbation theory. Finally, we propose a method to achieve the quantum simulation of the quantum Rabi model with a Stark term by using the internal and vibrational degrees of freedom of a trapped ion, and demonstrate its performance with numerical simulations considering realistic physical parameters.
\end{abstract}
\maketitle

\section{Introduction}
Understanding the interactions that emerge among two-level atoms (qubits) and bosonic field modes is of major importance for the development of quantum technologies. The qubit-boson interaction governs the dynamics of distinct quantum platforms such as cavity QED~\cite{Raimond01}, trapped ions~\cite{Leibfried03} or superconducting circuits~\cite{Clarke08}, that can achieve the so-called strong coupling (SC) regime. Here, the qubit-boson Rabi coupling $g$ is much smaller than the field frequency, but it is larger than the coupling to the environment. In these conditions, the Jaynes-Cummings (JC) model~\cite{Jaynes63} that appears after applying the rotating-wave approximation (RWA) provides an excellent description of the system. On resonance, the frequency of the bosonic mode $\omega$ equals the frequency of the qubit $\omega_0$ and the JC model predicts a coherent exchange of a single energy excitation between the atom and the field leading to Rabi oscillations. In the JC model, these Rabi oscillations are restricted to pairs of states known as JC doublets. When the qubit-boson coupling increases and reaches the ultrastrong coupling (USC)~\cite{Gunter09,Niemczyk10,Rossatto17,Kockum19,Forn19} regime ($g/\omega\gtrsim0.1$) or the deep-strong coupling (DSC)~\cite{Yoshihara17} regime ($g/\omega\gtrsim1$), the RWA does not hold and the full quantum Rabi model (QRM) has to be considered~\cite{Rabi36,Braak16}. Recently, an exact analytical solution for the QRM was proposed~\cite{Braak11}. Unlike the JC model, the QRM dynamics does not show clear features, until it reaches the DSC regime, where periodic collapses and revivals of the qubit initial-state survival probability are predicted~\cite{Casanova10}. 

The QRM with a Stark coupling term, named the Rabi-Stark model~\cite{Eckle17} (in the following we will also use that denomination) was first considered by Grimsmo and Parkins~\cite{Grimsmo13,Grimsmo14}. On the one hand, the study of its energy spectrum~\cite{Eckle17,Maciejewski14,Xie19,Chen2020} has revealed some interesting features such as a spectral collapse or a first-order phase transition~\cite{Xie19}, which connects it with the two-photon~\cite{Travenec12,Felicetti18_1,Felicetti18_2,Cong19,Xie17,Li2019} or anisotropic~\cite{Xie14,Xie2019} QRMs. On the other hand, dynamical features of the JC model with a Stark coupling term have been studied in the past~\cite{Pellizzari94,Solano00,Franca01,Solano05,Franca05,Prado13}. The Stark coupling is useful to restrict the resonance condition and the Rabi oscillations to a preselected JC doublet, leaving the other doublets in a dispersive regime. This selectivity has found applications for state preparation and reconstruction of the bosonic modes in cavity QED~\cite{Pellizzari94,Franca01} or trapped ions~\cite{Solano00,Solano05,Franca05}. In light of the above, the dynamical study of the full QRM with a Stark coupling term in the SC and USC regimes is well justified.

In this article, we study the dynamical behaviour of the QRM with a Stark term, i.e. the Rabi-Stark model, and show that the interplay between the Stark and Rabi couplings gives rise to selective $k$-photon interactions in the SC and USC regimes. Note that, previously, $k$-photon (or multiphoton) resonances have been investigated in the linear QRM~\cite{Ma15,Garziano15}, driven linear qubit-boson couplings~\cite{Nha00,Chough00,Klimov04,Casanova18,Puebla19_1,Puebla19_2} or nonlinear couplings~\cite{Shore93,Vogel95,Cheng18}, and recently have found applications for quantum-information science~\cite{Macri18,Boas19}. In our case, $k$-photon transitions appear as higher-order processes of the linear QRM, while the Stark coupling is responsible for the selective nature of these interactions. Using time-dependent perturbation theory we characterise these $k$-photon interactions, whose strength scales as $(g/\omega)^k$. Moreover, we design a method to simulate the Rabi-Stark model in a wide parameter regime using a single trapped ion. We validate our proposal with numerical simulations which show an excellent agreement between the dynamics of the Rabi-Stark model and the one achieved by the trapped-ion simulator.

\section{Model}
The Hamiltonian of the Rabi-Stark model is
\begin{equation}\label{QRS1}
H=\frac{\omega_0}{2}\sigma_z +\omega a^\dag a + \gamma a^\dag a \sigma_z + g(\sigma_+ +\sigma_-)(a+a^\dag)
\end{equation}
where $\omega_0$ is the frequency of the qubit or two-level system, $\omega$ is the frequency of the bosonic field, and $\gamma$ and $g$ are the couplings of the Stark and Rabi terms, respectively. Note that the Stark term is diagonal in the bare basis $\big\{|{\rm e}\rangle,|{\rm g}\rangle\big\}\otimes|n\rangle$ (where $\sigma_z|{\rm e}\rangle=|{\rm e}\rangle$, $\sigma_z|{\rm g}\rangle=-|{\rm g}\rangle$ and $a^\dagger a|n\rangle=n|n\rangle$), and it can be interpreted as a qubit energy shift that depends on the bosonic state. If we move to an interaction picture with respect to (w.r.t.) the first three terms in Eq.~(\ref{QRS1}), the system Hamiltonian reads (see Appendix A for additional details)
\begin{equation}\label{QRSIntPic}
H_I(t)=\sum_{n=0}^{\infty}\Omega_n(\sigma_+ e^{i\delta^+_{n} t}+\sigma_- e^{i\delta^-_{n} t})|n\!+\!1 \rangle\langle n| + {\rm H.c.}
\end{equation}
where $\Omega_n=g\sqrt{n+1}$, $\delta_n^+=\omega+\omega^0_n$ and $\delta_n^-=\omega-\omega_n^0$, with $\omega_n^0=\omega_0+\gamma(2 n +1)$. If $\gamma=0$, these detunings are independent of the state $n$, and, for $|\delta^+| \gg \Omega_n$ and $\delta^-=\omega-\omega_0=0$ ($|\delta^-| \gg \Omega_n$ and $\delta^+=\omega+\omega_0=0$), a resonant JC (anti-JC) Hamiltonian is recovered when fast-rotating terms are averaged out by invoking the RWA. In these conditions, the dynamics leads to Rabi oscillations between the states $|{\rm e},n\rangle \leftrightarrow |{\rm g},n+1\rangle$ ($|{\rm g},n\rangle \leftrightarrow |{\rm e},n+1\rangle$) for every $n$, and at a rate proportional to $\Omega_n$. These interactions are not selective as they apply to all Fock states in the same manner. 

\subsection{Selectivity in one-photon interactions}
\begin{figure}
\centering
\includegraphics[width=1\linewidth]{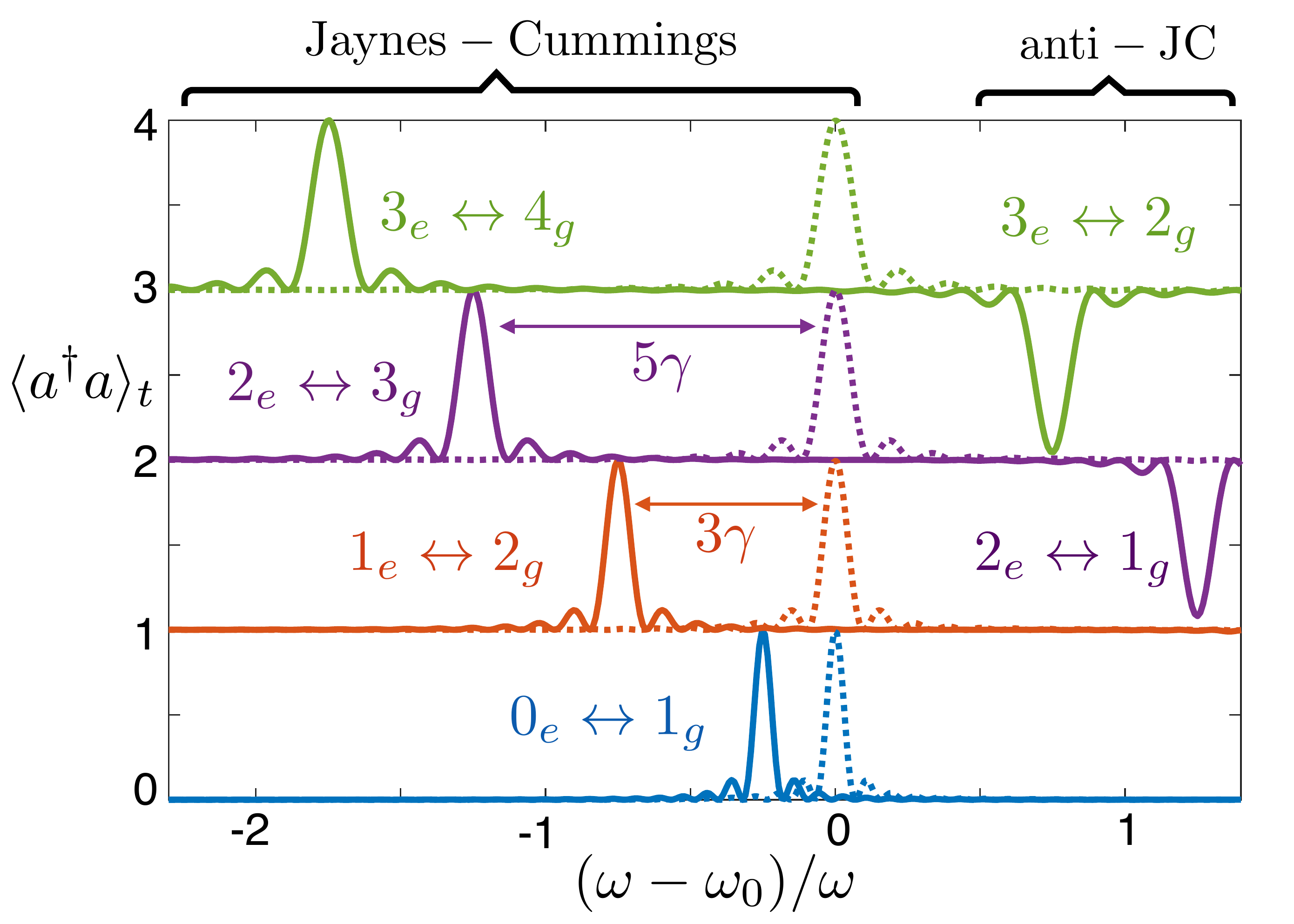}
\caption{One-photon selective interactions of the Rabi-Stark model. Hamiltonian (\ref{QRS1}) acts during a time $t=\pi/2\Omega_n$ and we calculate $\langle a^\dagger a\rangle$ for different ratios of $\omega_0/\omega$ and initial states $|{\rm e},0\rangle$ (blue), $|{\rm e},1\rangle$ (orange), $|{\rm e},2\rangle$ (purple) and $|{\rm e},3\rangle$ (green) with fixed couplings $\gamma/\omega=-0.25$ and $g/\omega=0.02$ (solid lines). If $\gamma=0$, all JC peaks would be at $\omega-\omega_0=0$ (dashed lines).}\label{Fig1}
\end{figure}

The presence of a nonzero Stark coupling $\gamma$ makes these detunings dependent on $n$, allowing to identify a resonance condition for a selected Fock state $n=N_0$, while the rest of Fock states stay out of resonance. From Eq.~(\ref{QRSIntPic}) we note that if $\delta^{-}_{N_0}=0$ ($\delta^{+}_{N_0}=0$) and $|\delta^{-}_{n\neq N_0}|\gg \Omega_{n\neq N_0}$ ($|\delta^{+}_{n\neq N_0}|\gg \Omega_{n\neq N_0}$), the dynamics of Hamiltonian (\ref{QRS1}) will produce a resonant one-photon JC (anti-JC) interaction only in the subspace $\{|{\rm e}\rangle,|{\rm g}\rangle\}\otimes\{|N_0\rangle,|N_0+1\rangle\}$.  This is observed in Fig.~\ref{Fig1}, where resonance peaks appear for initial states $|{\rm e},n\rangle$ with different number $n$. Here a one-photon Rabi oscillation occurs if $\omega-\omega_0=\gamma(2n+1)$ i.e. $\delta^{-}_{n}=0$. In Fig.~\ref{Fig1}, we vary $(\omega-\omega_0)/\omega$ in the $x$ axis for fixed $\gamma/\omega=-0.25$ and $g/\omega=0.02$, and meet this resonance condition for $n=0,1,2,3$ that correspond to the four peaks on the left side (solid lines). The other two peaks on the right correspond to $\delta^{+}_{n}=0$ resonances leading to one-photon anti-JC interactions for $n=1,2$.

%b) Time evolution of observables $\langle\sigma_z\rangle$ and $\langle a^\dag a\rangle$ for initial state $|\!\!\uparrow,1\rangle$ and resonance conditions that correspond to the first peak, $\omega_0+\omega=-\gamma$. c) Time evolution of $\langle\sigma_z\rangle$ and $\langle a^\dag a\rangle$ for initial state $|\!\!\uparrow,0\rangle$ and resonance conditions that correspond to the fourth peak, $\omega_0-\omega=\gamma$.}

\subsection{Multi-photon interactions}
As revealed in the Introduction, besides one-photon transitions, the Rabi-Stark Hamiltonian produces selective $k$-photon interactions. The characterization of these interactions is the main result of our work. Unlike the selective one-photon interactions, which appear due to the interplay between the Stark term and the rotating or counter-rotating terms, these selective multi-photon interactions are a direct consequence of the interplay between the Stark term and both the rotating and counter-rotating terms. Calculating the Dyson series for Eq.~(\ref{QRSIntPic}), we obtain that the second-order Hamiltonian is 
\begin{equation}\label{SecondOrder1}
H_I^{(2)}=\sum_{n=0}^{\infty}\big(\Delta_n^{\rm e}\sigma_+\sigma_- +\Delta_n^{\rm g}\sigma_-\sigma_+ \big)|n\rangle\langle n|
\end{equation}
where $\Delta_n^{\rm e}=\Omega^2_{n-1}/\delta^+_{n-1}-\Omega^2_{n}/\delta_n^-$ and $\Delta_n^{\rm g}=\Omega^2_{n-1}/\delta^-_{n-1}-\Omega^2_n/\delta_n^+$, plus a time-dependent part oscillating with frequencies $\delta_{n+1}^++\delta_n^-=2\omega+2\gamma$, $\delta_{n+1}^-+\delta_n^+=2\omega-2\gamma$, and $\delta_{n}^\pm,\delta_{n+1}^\pm$ that is averaged out due to the RWA (see Appendix A for a detailed derivation). 

The third-order Hamiltonian leads to three-photon transitions described by the following Hamiltonian~(see Appendix A for a complete derivation)
\begin{equation}\label{ThirdOrder1}
H_I^{(3)}(t)=\sum_{n=0}^{\infty}\big(\Omega^{(3)}_{n+}e^{i\delta^{(3)}_{n+}t}\sigma_++\Omega^{(3)}_{n-}e^{i\delta^{(3)}_{n-}t}\sigma_-\big)|n\!+\!3\rangle \langle n| + {\rm H.c.},
\end{equation}
where $\Omega^{(3)}_{n\pm}=g^3\sqrt{(n+3)!/n!}/2\delta^\pm_n(\omega\mp\gamma)$ and $\delta^{(3)}_{n\pm}=\delta^\pm_{n+2}+\delta^\mp_{n+1}+\delta_n^\pm=2\omega+\delta_{n+1}^\pm$. According to this, a JC type three-photon process occurs for $|{\rm e},N_0\rangle$ if $\delta^{(3)}_{N_0-}=0$ producing population exchange between the states $|{\rm e}, N_0\rangle\leftrightarrow|{\rm g}, N_0+3\rangle$. For the state $|{\rm g}, N_0\rangle$, anti-JC-type transitions to the state $|{\rm e}, N_0+3\rangle$ occur when $\delta^{(3)}_{N_0+}=0$. In the following we will check the validity of these effective Hamiltonians by numerically calculating the dynamics of Hamiltonian~(\ref{QRS1}).

\begin{figure}
\centering
\includegraphics[width=1\linewidth]{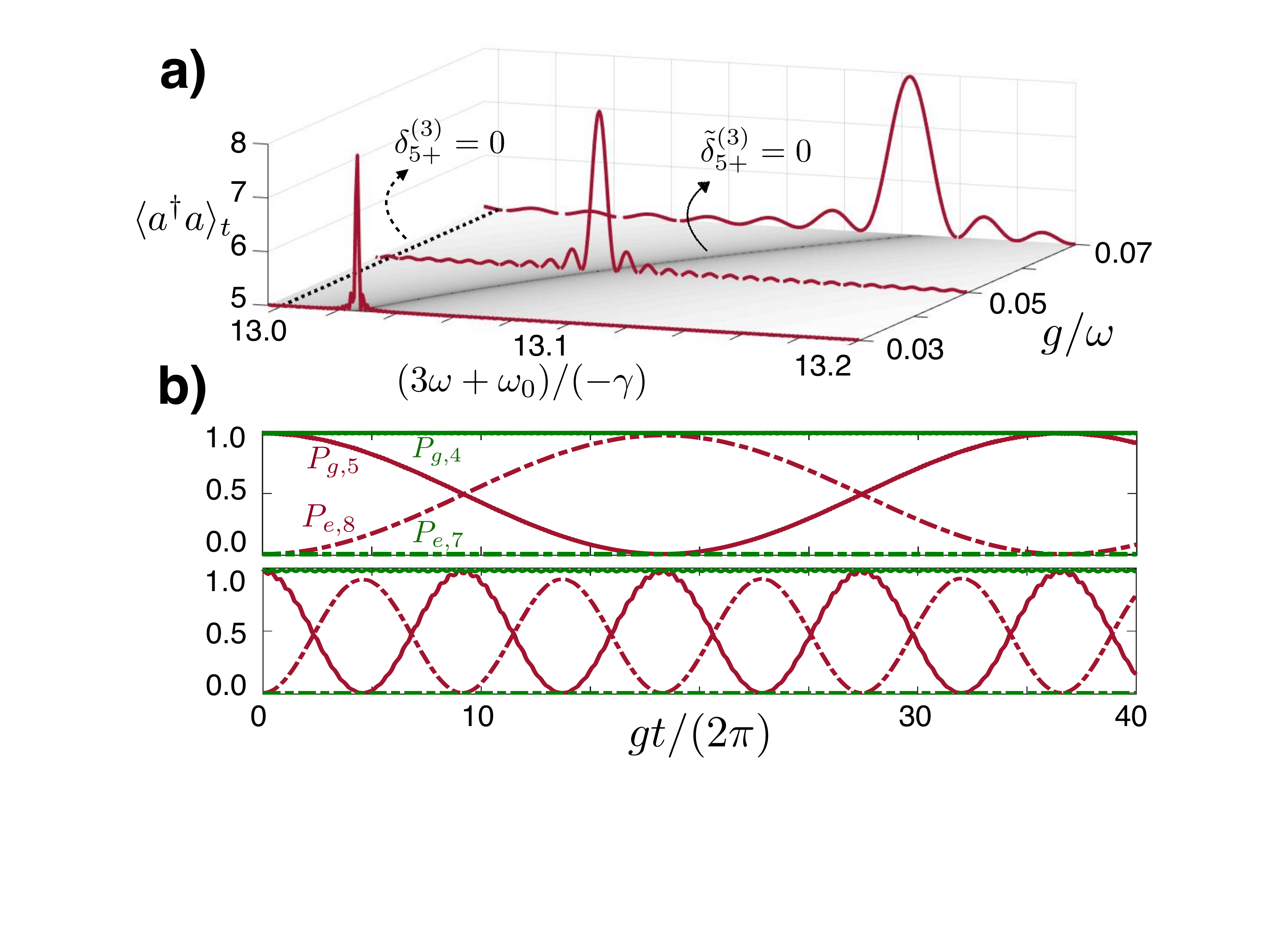}
\caption{Three-photon selective interactions of the Rabi-Stark model. (a) Resonance spectrum of anti-JC-like three-photon process for state $|{\rm g},5\rangle$. After a time $t=\pi/2\Omega^{(3)}_{5+}$, $\langle a^\dagger a\rangle$ is shown for $\gamma/\omega=-0.4$. The peaks appear shifted from $\delta_{5+}^{(3)}=0$ (dashed line) at $\tilde{\delta}_{5+}^{(3)}=0$ which corresponds to the dark curve in the XY plane representing the lower values of $\log_{10}{|\delta^{(3)}_{5+}|}$. (b) Time evolution of populations $P_{{\rm g},4}$ (solid) and $P_{{\rm e},7}$ (dashed) for initial state $|{\rm g},4\rangle$ (green) and populations $P_{{\rm g},5}$ and $P_{{\rm e},8}$ for initial state $|{\rm g},5\rangle$ (red) for $g/\omega=0.05$ (up) and $g/\omega=0.1$ (down).}
\label{Fig2}
\end{figure}

In Fig.~\ref{Fig2}(a) we let the system evolve for a time $t=\pi/2\Omega^{(3)}_{5+}$ for a fixed value of $\gamma/\omega=-0.4$ and calculate the average number of photons $\langle a^\dagger a\rangle$ for different values of $\omega_0/\omega$ and couplings $g/\omega$. We do this for the initial state $|{\rm g},N_0=5\rangle$, near the resonance point $\delta^{(3)}_{5+}=3\omega+\omega_0+13\gamma=0$. We observe that resonances do not appear when $\delta^{(3)}_{5+}=0$, see dashed line on the left, owing to a resonance-frequency shift that depends on the value of $g$. To explain this we go to an interaction picture w.r.t. Eq.~(\ref{SecondOrder1}), then, the oscillation frequencies in Eq.~(\ref{ThirdOrder1}) will be shifted to $\tilde{\delta}^{(3)}_{n+}=\delta^{(3)}_{n+}+\Delta^e_{n+3}-\Delta_n^g$ and $\tilde{\delta}^{(3)}_{n-}=\delta^{(3)}_{n-}+\Delta^g_{n+3}-\Delta_n^e$. In the XY plane of Fig.~\ref{Fig2}(a) we make a grayscale colour plot of $\log_{10}{|\tilde{\delta}^{(3)}_{5+}|}$ as a function of $\omega_0$ and $g$ and see that the minima of $\tilde{\delta}^{(3)}_{5+}$ (dark line) is in very good agreement with the point in which the three-photon resonance appears (the logarithm scale is used to better distinguish the zeros of $\tilde{\delta}^{(3)}_{5+}$). 

To show that the three-photon interaction applies only to the preselected subspace, in Fig.~\ref{Fig2}(b) we plot the evolution of initial states $|{\rm g},4\rangle$ and $|{\rm g},5\rangle$. As expected, the last term exchanges population with the state $|{\rm e},8\rangle$ while the other remains constant. In addition, for $g/\omega=0.05$ (upper figure), the transition is slower but most of the population is transferred to $|{\rm e},8\rangle$ at time $t=\pi/2\Omega^{(3)}_{5+}$. For $g/\omega=0.1$ (lower figure) the exchange rate is much faster but the transfer is not so efficient. 

In this context, higher-order selective interactions will be produced by the Rabi-Stark model and could in principle be tracked by the calculation of higher-order Hamiltonians. However, being a high-order process, its strength decreases with order $k$ since $\Omega^{(k)}/\omega\propto (g/\omega)^k$. Then, high-order processes require longer times to be observed which may exceed the decoherence times of the system. See Appendix B for a numerical analysis of dissipative effects. In any case, we find interesting to study the case for a higher $k$. Following the same procedure as for calculating Eqs.~(\ref{SecondOrder1}) and (\ref{ThirdOrder1}), we conclude that for even $k$, the $k$-th order Hamiltonian will not produce selective interactions as they will average out as a consequence of the RWA. For odd $k$, the $k$-th order Hamiltonian predicts a $k$-photon transition of the form
\begin{equation}\label{kthOrder1}
H_I^{(k)}(t)=\sum_{n=0}^{\infty}\big(\Omega^{(k)}_{n+}e^{i\delta^{(k)}_{n+}t}\sigma_++\Omega^{(k)}_{n-}e^{i\delta^{(k)}_{n-}t}\sigma_-\big)|n\!+\!k\rangle\langle n| + {\rm H.c.},
\end{equation}
where 
\begin{equation}\label{det}
\delta^{(k)}_{n\pm}=\sum_{s=0}^{k-1} \delta_{n+s}^\pm+\delta_{n+s+1}^\mp+\delta^\pm_{k}=(k-1)\omega+\delta^{\pm}_{n+(k-1)/2}
\end{equation}
and 
\begin{equation}\label{RabiFreq}
\Omega^{(k)}_{n\pm}=\frac{g^k}{(k-1)!!(\omega\mp\gamma)^{\frac{k-1}{2}}}\sqrt{\frac{(n+k)!}{n!}}\prod^{k-2}_{s=1,3...}\frac{1}{\delta^{(s)}_{n\pm}}.
\end{equation}

To gain some physical intuition about the difference between odd and even orders, one can consider the symmetry in Hamiltonian~(\ref{QRS1}) \cite{Eckle17,Xie19}. As in the QRM \cite{Casanova10}, due to parity symmetry, transitions between $|{\rm e},n\rangle$ and $|{\rm g},n+k\rangle$ states are not allowed for even $k$. On the contrary, for odd $k$, transitions between these states are possible when the energy cost of an atomic excitation is similar to that of $k$ bosons, i.e. $\omega_0\approx k\omega$.

Using Eqs.~(\ref{det}) and (\ref{RabiFreq}) and with the help of numerical simulations, it is easy to find $k$-photon processes to validate the effective Hamiltonian (\ref{kthOrder1}). Here, numerical simulations are required as the analytic calculation of the exact resonance frequencies of higher-order processes rapidly becomes challenging. For example, for tracking a JC-type five-photon interaction for $N_0$, we use the condition $\delta^{(5)}_{N_0-}=0$ to retrieve an approximate value for the qubit frequency of $\omega^c_0= 5\omega-\gamma(2N_0+5)$. Then, we calculate the time evolution governed by Hamiltonian~(\ref{QRS1}) for a time $t=\pi/2\Omega^{(5)}_{N_0-}$ and plot $\langle\sigma_+\sigma_-\rangle$ for different values of $\omega_0$ close to $\omega_0^c$ until we find a peak corresponding to the resonant five-photon interaction.

\begin{figure}
\centering
\includegraphics[width=1\linewidth]{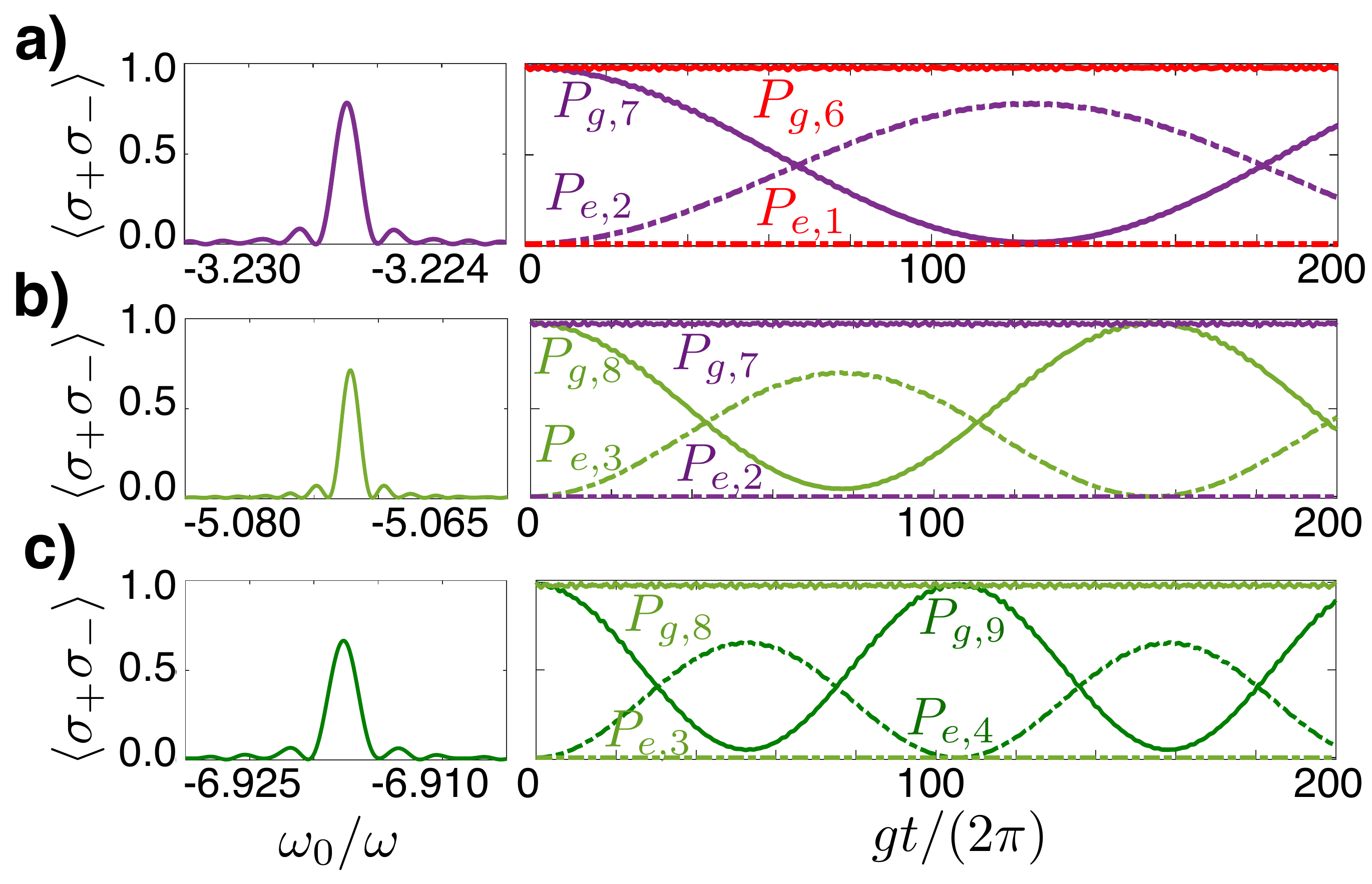}
\caption{Five-photon selective interactions of the Rabi-Stark model. a) On the left, $\langle\sigma_+\sigma_-\rangle$ is shown after a time $t=\pi/2\Omega^{(5)}_{2-}$, for different values of $\omega_0/\omega$ around $\omega_0^c=5\omega-\gamma(2\times2+5)$ and initial state $|{\rm g},7\rangle$. Here, $g/\omega=0.1$ and $\gamma/\omega=0.9$. On the right, time evolution of populations $P_{{\rm e},2}$ and  $P_{{\rm g},7}$ for initial state $|{\rm g},7\rangle$ and populations $P_{{\rm e},1}$ and $P_{{\rm g},6}$ for initial state $|{\rm g},8\rangle$, for $\omega_0/\omega=-3.227$. b) and c) The same procedure with initial state $|{\rm g},8\rangle$ and $|{\rm g},9\rangle$, where the peaks appear for $\omega_0/\omega=-5.072$ and $\omega_0/\omega=-6.918$. } 
\label{Fig3}
\end{figure}

As an example, in Fig.~\ref{Fig3} we show these resonances for $N_0=2$, $3$ and $4$, with $g/\omega=0.1$ and $\gamma/\omega=0.9$. We find resonance peaks for $\omega_0/\omega=-3.227$,$-5.072$ and $-6.918$ which are close to the ones obtained with the approximate formula $\omega_0^c/\omega=-3.1$,$-4.9$ and $-6.7$. In comparison with the three-photon processes, five-photon transitions are slower, and the population transfer to the preselected state is partial for $g/\omega=0.1$. It is interesting to note that the revival of the initial state as well as the selectivity condition are maintained at the beginning of the USC regime. Note that for $\omega_0/\omega=-3.227$, an exchange between states $|{\rm g},7\rangle \leftrightarrow |{\rm e},2\rangle$ occurs while the neighboring states $ |{\rm g},6\rangle$ and  $|{\rm e},1\rangle$ are completely out of resonance. In this respect, with larger coupling constants such as $g/\omega\approx0.3$ one would still get signatures of selectivity, but the interaction will not longer be a JC (or anti-JC) type $k$-photon interaction as it would involve states out of the selected JC (or anti-JC) doublet. In Fig.~\ref{Fig3} the population transfer from $|{\rm g},N_0+5\rangle$ to $|{\rm e},N_0\rangle$ is already partial, and interestingly, the remaining population goes to states $|{\rm g},N_0+1\rangle$ and $|{\rm g},N_0-1\rangle$.

To experimentally verify our predictions regarding the selective $k$-photon interactions of the Rabi-Stark model, in the next chapter we propose an experimental implementation of the model.

\section{Implementation with trapped ions}
Trapped ions are excellent quantum simulators~\cite{Leibfried03,Blatt12}, with experiments for the one-photon QRM~\cite{Pedernales15,Puebla16,Lv18} and proposals for the two-photon QRM~\cite{Felicetti15,Puebla17}. In the following, we propose a route to simulate the Rabi-Stark model using a single trapped ion. 

The Hamiltonian of a single trapped ion interacting with co-propagating laser beams labeled with $j$ can be written, in an interaction picture w.r.t. the free energy Hamiltonian $H_0=\omega_I/2\sigma_z+\nu a^\dagger a$, as~\cite{Leibfried03}  
\begin{equation}\label{TIHamil}
H=\sum_{j}\frac{\Omega_j}{2} \sigma^+e^{i\eta(ae^{-i\nu t}+a^\dag e^{i\nu t})}e^{-i(\omega_j-\omega_I)t}e^{i\phi_j} +{\rm H.c.}
\end{equation}
Here $\Omega_j$ is the Rabi frequency, $\eta$ and $a^\dagger (a)$ are the Lamb-Dicke (LD) parameter and the creation (annihilation) operator acting on vibrational phonons, $\nu$ is the trap frequency, $\omega_j-\omega_I$ is the detuning of the laser frequency $\omega_j$ w.r.t. the carrier frequency $\omega_I$, and $\phi_j$ accounts for the phase of the laser. 

As a possible implementation of the Rabi-Stark model we consider two drivings acting near the first red and blue sidebands, and a third one on resonance with the carrier interaction $\omega_{\rm S}=\omega_0$. The Hamiltonian in the LD regime, $\eta \sqrt{\langle n\rangle}\ll 1$, and after the vibrational RWA, reads
\begin{equation}\label{Scheme2}
H_{\rm LD}=-ig_ra \sigma^+ e^{-i\delta_rt}  -ig_b a^\dagger \sigma^+ e^{-i\delta_bt} -  g_{\rm S}\sigma^++{\rm H.c.}
\end{equation}
where $\omega_{r,b}=\omega_0\mp \nu +\delta_{r,b}$,  $g_{r,b}=\eta\Omega_{r,b}/2$, $\phi_{r,b,{\rm S}}=-\pi$ and $g_{\rm S}=\frac{\Omega_{\rm S}}{2}(1-\eta^2/2)-\frac{\Omega_{\rm S}}{2}\eta^2a^\dag a=\frac{\Omega_0}{2}-\gamma a^\dagger a$. Dependence of the carrier interaction on the phonon number appears when considering the expansion of $e^{i\eta(a+a^\dag)}$ up to the second order in $\eta$. At this point, if $\delta_r=-\delta_b=\omega^{\rm R}$, Eq.~(\ref{Scheme2}) can already be mapped to a Rabi-Stark model in a frame rotated by $-\omega^{\rm R}a^\dagger a$. However, the engineered Hamiltonian cannot explore all regimes of the model, as $\Omega_0$ and $\gamma$ cannot be independently tuned, thus restricting the Hamiltonian to regimes where $\gamma\ll\Omega_0$. This issue can be solved by moving to an interaction picture w.r.t. $\frac{\Omega_{\rm DD}}{2}\sigma_x - \omega^{\rm R}a^\dagger a$, where $\Omega_{\rm DD}=-(\Omega_0+\omega_0^{\rm R})$, and by shifting the detunings by $\delta_{r,b}=\Omega_{\rm DD}\pm\omega^{\rm R}$. The resulting Hamiltonian, after ignoring terms oscillating at $\Omega_{\rm DD}$, is 
\begin{equation}\label{TIQRS}
H^{II}_{\rm LD}=\frac{\omega_0^{\rm R}}{2}\sigma_x+\omega^{\rm R} a^\dagger a + g \sigma_y(a+a^\dagger)+\gamma a^\dagger a\sigma_x.
\end{equation}
Here $g=(\eta\Omega_r/4)(1-\epsilon_{\rm S})$ if $\Omega_b=\Omega_r(1-\epsilon_{\rm S})/(1+\epsilon_{\rm S})$ with $\epsilon_{\rm S}=\Omega_{\rm S}/\nu$. See Appendix C for a detailed derivation of Eq.~(\ref{TIQRS}). Notice that Eqs.~(\ref{QRS1}) and (\ref{TIQRS}) are equivalent by simply changing the qubit basis. For Eq.~(\ref{TIQRS}), the diagonal basis is given by $\{|+\rangle,|-\rangle\}\otimes|n\rangle$, where $\sigma_x|\pm\rangle=\pm|\pm\rangle$. The parameters of the model are now $\omega^{\rm R}_0=-(\Omega_0+\Omega_{\rm DD})$, $\omega^{\rm R}=(\delta_r-\delta_b)/2$, and $\gamma=\eta^2\Omega_{\rm S}/2$. Regimes where $\gamma<0$ can be also reached by taking $\phi_{\rm S}=0$, however, the frequency of the rotating frame changes to $\Omega_{\rm DD}=\Omega_0-\omega_0^{\rm R}$. Moreover, in this case $g=(\eta\Omega_r/4)(1+\epsilon_{\rm S})$ if $\Omega_b=\Omega_r(1+\epsilon_{\rm S})/(1-\epsilon_{\rm S})$. 

Regarding the initial-state preparation, laser-cooling techniques can be used to initialize the system in state $|\rm g,0\rangle$ with high fidelity. Later, the carrier interaction can used to prepare an arbitrary qubit state, and this can be combined with JC or anti-JC interactions (using the first-red or first-blue sidebands), to prepare an arbitrary Fock state~\cite{Meekhof96}. In addition, high-$n$ Fock states can be prepared combining controlled depolarizing noise applied to the ion internal state with anti-JC interactions beyond the LD regime~\cite{Cheng18}. Finally, population of the state $|\rm g\rangle$ can be measured via resonance-fluorescence detection. This is then combined with the Fourier cosine transform to extract the populations of each Fock state~\cite{Meekhof96}.

\begin{figure}
\centering
\includegraphics[width=1\linewidth]{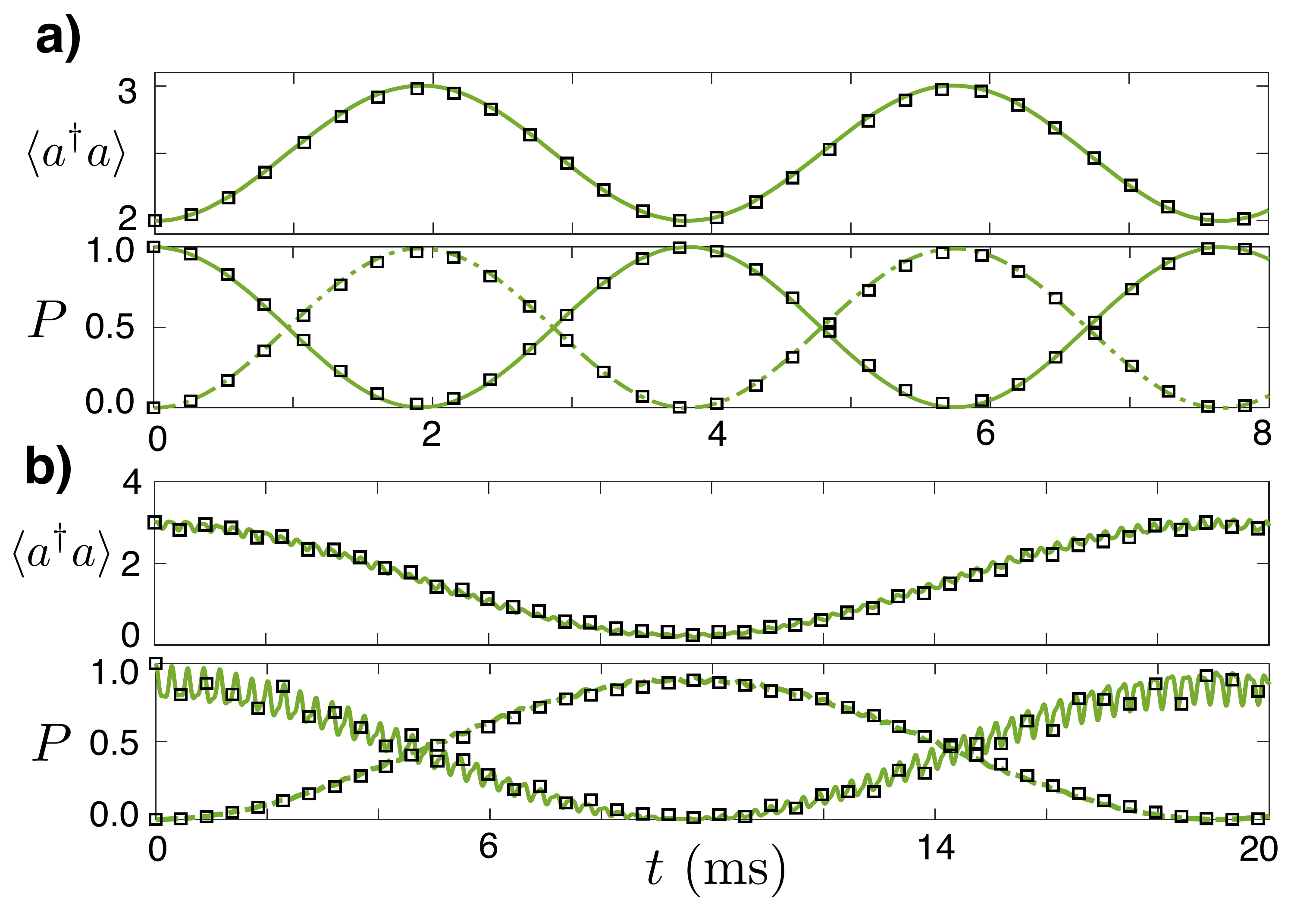}
\caption{Selective one-photon and three-photon interactions with a trapped ion. a) Time evolution of the mean number of phonons and populations $P_{+,2}$ (solid) and $P_{-,3}$ (dashed) starting from state $|+,2\rangle$ for $g/\omega^{\rm R}=0.05$, $\gamma/\omega^{\rm R}=-0.4$ and $\omega_0^{\rm R}/\omega^{\rm R}=3$. b) Time evolution of the mean number of phonons and populations $P_{+,3}$ (solid) and $P_{-,0}$ (dashed) starting from state $|+,3\rangle$ for $g/\omega^{\rm R}=0.3$, $\gamma/\omega^{\rm R}=-0.1$ and $\omega_0^{\rm R}/\omega^{\rm R}=-2.4385$. Solid green lines evolve according to Eq.~(\ref{TIQRS}) while black squares evolve according to Eq.~(\ref{TIHamil}).} 
\label{Fig4}
\end{figure}

In the following, we verify the feasibility of the proposal by comparing the dynamics generated by the Hamiltonian in Eq.~(\ref{TIHamil}) with the one of the Rabi-Stark model at Eq.~(\ref{TIQRS}). The results are shown in Figs.~\ref{Fig4}(a) and (b) for one-photon and three-photon oscillations respectively. The experimental parameters we use in Fig.~\ref{Fig4}(a) are $\nu=(2\pi)\times4.98$ MHz for the trapping frequency, $\eta=0.1$ for the LD parameter and $\Omega_{\rm S}=(2\pi)\times120$ kHz for the carrier driving, leading to a Stark coupling of $|\gamma|=(2\pi)\times0.6$ kHz. We consider a Stark coupling of $\gamma/\omega^{\rm R}=-0.4$, a Rabi coupling of $g/\omega^{\rm R}=0.05$, and $\omega_0^{\rm R}=\omega^{\rm R}-\gamma(2N_0+1)$ with $N_0=2$. To achieve this regime the experimental parameters are $\Omega_r=(2\pi)\times2.94$ kHz, $\Omega_b=(2\pi)\times3.08$ kHz, and $\Omega_{\rm DD}=(2\pi)\times114.86$ kHz. We observe that with the initial state $|+,2\rangle$, there is an exchange of population with the state $|-,3\rangle$. 
In Fig.~\ref{Fig4}(b) we show that selective three-photon oscillations of the Rabi-Stark model can be observed in some milliseconds. Starting from $|+,3\rangle$, we can observe the coherent population exchange with state $|-,0\rangle$. Here, the LD parameter is $\eta=0.05$ and the parameters of the model are $\gamma/\omega^{\rm R}=-0.1$, $g/\omega^{\rm R}=0.3$ and $\omega_0^{\rm R}/\omega^{\rm R}=-2.4385$ for which we require $\Omega_r=(2\pi)\times35.2$ kHz, $\Omega_b=(2\pi)\times36.9$ kHz, and $\Omega_{\rm DD}=(2\pi)\times123.5$ kHz. Although in the previous case we focused on the Rabi-Stark model in the strong and ultrastrong coupling regimes, it is noteworthy to mention that our method is still valid for larger ratios of $g/\omega$. Thus, our method represents a simple and versatile route to simulate the Rabi-Stark model in all important parameter regimes.

\section{Conclusions}

We studied the dynamics of the QRM with a Stark term in the strong and ultrastrong coupling regimes and characterize the novel $k$-photon interactions that appear by using time-dependent perturbation theory.  Due to the Stark-coupling term, these $k$-photon interactions are selective, thus their resonance frequency depends on the state of the bosonic mode. Finally, and with the support of detailed numerical simulations, we propose an implementation of the Rabi-Stark model with a single trapped ion. The numerical simulations show an excellent agreement between the dynamics of the trapped-ion system and the Rabi-Stark model.

\section*{acknowledgements}

We thank Kihwan Kim for useful discussions regarding the experimental implementation. Authors acknowledge financial support from Spanish Government PGC2018-095113-B-I00 (MCIU/AEI/FEDER, UE), Basque Government IT986-16, as well as from QMiCS (820505) and OpenSuperQ (820363) of the EU Flagship on Quantum Technologies, and EU FET Open Grant Quromorphic. L. C. acknowledges support by the China Postdoctoral Science Foundation (No. 2019M651463). S. F. acknowledges support from the European Research Council (ERC-2016-STG-714870). J. C. acknowledges support by the Juan de la Cierva grant IJCI-2016-29681. I. A. acknowledges support to the Basque Government PhD grant PRE-2015-1-0394.

\section*{Appendix A: Dyson series of the Rabi-Stark Hamiltonian}
The Rabi-Stark Hamiltonian as written in Eq.~(\ref{QRS1}) of the main text is
\begin{equation}\label{SQRS1}
H=\frac{\omega_0}{2}\sigma_z +\omega a^\dag a + \gamma a^\dag a \sigma_z + g(\sigma_+ +\sigma_-)(a+a^\dag),
\end{equation}
where $\sigma_z,\sigma_+,\sigma_-$ are operators of the two-level system and $a^\dagger$ and $a$ are infinite-dimensional creation and annihilation operators of the bosonic field. Using the ket-bra notation, the two-level matrices are $\sigma_+=|e \rangle\langle g|$, $\sigma_-=|g \rangle\langle e|$ and $\sigma_z=|e \rangle\langle e|-|g \rangle\langle g|$ where $|e\rangle$ and $|g\rangle$ are the excited and ground states of the two-level system, respectively. On the other hand, the bosonic operators can be written as
\begin{eqnarray}\label{SBosOp}
a^\dagger=\sum_{n=0}^{\infty}\sqrt{n+1}|n+1\rangle\langle n|,   \\
a=\sum_{n=0}^{\infty}\sqrt{n+1}|n\rangle\langle n+1|, 
\end{eqnarray}
where $|n\rangle$ is the $n$-th Fock state. With this notation, the Hamiltonian in Eq.~(\ref{SQRS1}) can be rewritten as 
\begin{eqnarray}\label{SQRS2}
H&=&\sum_{n=0}^{\infty}\omega_n^e|e \rangle\langle e|\otimes |n \rangle\langle n|+\omega^g_{n}|g \rangle\langle g|\otimes |n \rangle\langle n| \nonumber \\ &+&\Omega_n(|e \rangle\langle g|+|g \rangle\langle e|)\otimes(|n+1\rangle\langle n| +|n\rangle\langle n+1|), 
\end{eqnarray}
where $\omega_n^e=(\omega  +\gamma )n+\omega_0/2$, $\omega_n^g=(\omega  -\gamma )n-\omega_0/2$, and $\Omega_n=g\sqrt{n+1}$. We can move to an interaction picture w.r.t. the diagonal part of Eq.~(\ref{SQRS2}), and the non-diagonal elements will rotate as
\begin{eqnarray}\label{SNonDiag2}
|e \rangle\langle g|\otimes |n+1 \rangle\langle n| &\rightarrow& |e \rangle\langle g| \otimes |n+1 \rangle\langle n| e^{i(\omega^e_{n+1}-\omega_n^g)t}, \\
|g \rangle\langle e|\otimes |n+1 \rangle\langle n|  &\rightarrow& |g \rangle\langle e| \otimes |n+1 \rangle\langle n| e^{-i(\omega^e_n-\omega_{n+1}^g)t}, \\
|e \rangle\langle g|\otimes |n \rangle\langle n+1| &\rightarrow& |e \rangle\langle g| \otimes |n \rangle\langle n+1| e^{i(\omega^e_{n}-\omega_{n+1}^g)t}, \\
|g \rangle\langle e|\otimes |n \rangle\langle n+1|  &\rightarrow& |g \rangle\langle e| \otimes |n \rangle\langle n+1| e^{-i(\omega^e_{n+1}-\omega_{n}^g)t}, 
\end{eqnarray}
where $\delta^+_{n}=\omega_{n+1}^e-\omega_n^g=\omega+[\omega_0+\gamma(2n+1)]$ and $\delta^-_{n}=\omega_{n+1}^g-\omega_{n}^e=\omega-[\omega_0+\gamma(2n+1)]$. The Hamiltonian in the interaction picture can be then rewritten as
\begin{eqnarray}\label{SQRSIntPic}
H_I(t)&=&\sum_{n=0}^{\infty}\Omega_n(\sigma_+ e^{i\delta^+_{n} t}+\sigma_- e^{i\delta^-_{n} t})\otimes |n+1\rangle\langle n| \nonumber \\ &+& \Omega_n(\sigma_+ e^{-i\delta^-_{n} t}+\sigma_- e^{-i\delta^+_{n} t})\otimes |n \rangle\langle n+1|,
\end{eqnarray}
which corresponds to Eq.~(\ref{QRSIntPic}). 

\subsection{Second-order Hamiltonian}
The second-order Hamiltonian that corresponds to Eq.~(\ref{SQRSIntPic}) is given by~\cite{Sakurai94}
\begin{equation}\label{S2Dyson}
H^{(2)}(t)=-i\int_0^t dt' H_I(t)H_I(t').
\end{equation}
We can write $H_{I}(t)$ as
\begin{equation}\label{SredHI}
H_I(t)=\sum_{n=0}^{\infty} \Omega_n \Big(S_{n}(t)|n+1\rangle\langle n| +S^\dagger_{n}(t)|n\rangle\langle n+1|\Big),
\end{equation}
and, then,

\begin{widetext}
\begin{equation}\label{S2orderQRS}
H^{(2)}=-i\sum_{n,n'}\Omega_{n}\Omega_{n'}\Big(S_{n}(t)|n+1\rangle\langle n| +S^\dagger_{n}(t)|n\rangle\langle n+1|\Big)\int_0^tdt'\Big(S_{n'}(t')|n'+1\rangle\langle n'| +S^\dagger_{n'}(t')|n'\rangle\langle n'+1|\Big),
\end{equation}
which gives $H^{(2)}=H_A^{(2)}+H_B^{(2)}$, where
\begin{equation}\label{S2orderQRS_2}
H^{(2)}_A(t)=-i\sum_{n}\Omega_{n}^2\Big(S_{n}(t)\int_0^tdt'S^\dagger_{n}(t')\Big)|n+1\rangle\langle n+1| +\Omega^2_{n}\Big(S^\dagger_{n}(t)\int_0^tdt'S_{n}(t')\Big)|n\rangle\langle n| 
\end{equation}
gives diagonal elements and
\begin{equation}\label{S2orderQRS_3}
H^{(2)}_B(t)=-i\sum_{n}\Omega_{n}\Omega_{n+1}\Big(S_{n+1}(t)\int_0^tdt'S_{n}(t')\Big)|n+2\rangle\langle n| + \Omega_{n}\Omega_{n+1}\Big(S^\dagger_{n}(t)\int_0^tdt'S^\dagger_{n+1}(t')\Big)|n\rangle\langle n+2| 
\end{equation}
is related with two-photon processes. Calculating the two-level operators we obtain 

\begin{eqnarray}\label{S2orderSpinOp}
S_{n}(t)\int_0^tdt'S^\dagger_{n}(t')&=&\frac{i}{\delta^+_n}\sigma_+\sigma_- + \frac{i}{\delta^-_n}\sigma_-\sigma_+ -\frac{i}{\delta^+_n}\sigma_+\sigma_-e^{i\delta^+_n t} - \frac{i}{\delta^-_n}\sigma_-\sigma_+e^{i\delta^-_n t} \\
S^\dagger_{n}(t)\int_0^tdt'S_{n}(t')&=&-\frac{i}{\delta^+_n}\sigma_-\sigma_+ - \frac{i}{\delta^-_n}\sigma_+\sigma_-+ \frac{i}{\delta^+_n}\sigma_-\sigma_+e^{-i\delta^+_n t} + \frac{i}{\delta^-_n}\sigma_+\sigma_-e^{-i\delta^-_n t} \\
S_{n+1}(t)\int_0^tdt'S_{n}(t')&=& -\frac{i}{\delta^-_{n}}\sigma_+\sigma_-(e^{i(\delta^+_{n+1} +\delta^-_{n})t} -e^{i\delta^+_{n+1}t})  - \frac{i}{\delta^+_{n}}\sigma_-\sigma_+(e^{i(\delta^-_{n+1} +\delta^+_{n})t} -e^{i\delta^-_{n+1}t})\\
S^\dagger_{n}(t)\int_0^tdt'S^\dagger_{n+1}(t')&=&  \frac{i}{\delta^-_{n+1}}\sigma_-\sigma_+(e^{-i(\delta^+_{n} +\delta^-_{n+1})t} -e^{-i\delta^+_{n}t})  + \frac{i}{\delta^+_{n+1}}\sigma_+\sigma_-(e^{-i(\delta^-_{n} +\delta^+_{n+1})t} -e^{i\delta^-_{n}t}).
\end{eqnarray}
We can ignore the terms oscillating with $\pm\delta^{\pm}_n$, as these frequencies correspond to resonances of the first-order Hamiltonian and one-photon processes. Keeping the other terms we have that
\begin{equation}\label{S2orderQRS_4}
H^{(2)}_A\approx\sum_{n}\Omega_{n}^2\Big(\frac{1}{\delta_n^+}\sigma_+\sigma_- +\frac{1}{\delta_n^-}\sigma_-\sigma_+\Big)|n+1\rangle\langle n+1| -\Omega^2_{n}\Big(\frac{1}{\delta^+_n}\sigma_-\sigma_+ + \frac{1}{\delta^-_n}\sigma_+\sigma_-\Big)|n\rangle\langle n| 
\end{equation}
and
\begin{eqnarray}\label{S2orderQRS_5}
H^{(2)}_B(t)\approx \sum_{n}-\Omega_{n}\Omega_{n+1}\Big(\frac{1}{\delta^-_{n}}\sigma_+\sigma_-e^{i(\delta^+_{n+1} +\delta^-_{n})t}   + \frac{1}{\delta^+_{n}}\sigma_-\sigma_+e^{i(\delta^-_{n+1} +\delta^+_{n})t} \Big)|n+2\rangle\langle n| \nonumber\\ 
+\Omega_{n}\Omega_{n+1}\Big(\frac{1}{\delta^-_{n+1}}\sigma_-\sigma_+e^{-i(\delta^+_{n} +\delta^-_{n+1})t}   + \frac{1}{\delta^+_{n+1}}\sigma_+\sigma_-e^{-i(\delta^-_{n} +\delta^+_{n+1})t} \Big)|n\rangle\langle n+2|. 
\end{eqnarray}

The two-photon transition terms in Eq.~(\ref{S2orderQRS_5}) oscillate with frequencies $\delta^+_n+\delta^-_{n+1}=2\omega-2\gamma$ and $\delta^+_{n+1}+\delta^-_{n}=2\omega+2\gamma$, which are zero only in the points of the spectral collapse. Thus, we do not expect to see two-photon transitions in the regime where the Hamiltonian is bounded from below, i.e. the ground-state energy is finite~\cite{Eckle17,Xie19}. The terms in Eq.~(\ref{S2orderQRS_4}) will induce an additional Stark shift that can induce a shift in the resonance conditions of the higher-order processes, as we will see later. The Hamiltonian can be simplified to
\begin{equation}\label{S2orderQRS_6}
H^{(2)}_A\approx\sum_{n}\Bigg\{\Big(\frac{\Omega_{n-1}^2}{\delta_{n-1}^+} -\frac{\Omega_{n}^2}{\delta_n^-}\Big)\sigma_+\sigma_- +\Big(\frac{\Omega_{n-1}^2}{\delta_{n-1}^-} -\frac{\Omega_{n}^2}{\delta_n^+}\Big)\sigma_-\sigma_+ \Bigg\} |n\rangle\langle n| 
\end{equation}

\subsection{Third-order Hamiltonian}
The third-order Hamiltonian is calculated by 
\begin{equation}\label{S3Dyson}
H^{(3)}(t)=(-i)^2\int_0^t dt' \int_0^{t'}dt'' H_I(t)H_I(t')H(t'').
\end{equation}

Following the same notation of the previous section, the third-order Hamiltonian is 
\begin{equation}\label{S3orderQRS}
H^{(3)}(t)=-\sum_{n, n', n''}\Omega_{n}\Omega_{n'}\Omega_{n''}\Big(S_{n}(t)|n+1\rangle\langle n| +{\rm H.c.}\Big)\int_0^t dt' \Big(S_{n'}(t')|n'+1\rangle\langle n'| +{\rm H.c.}\Big) \int_0^{t'}dt'' \Big(S_{n''}(t'')|n''+1\rangle\langle n''| +{\rm H.c.}\Big).
\end{equation}
If we focus on the three-photon resonances, the following Hamiltonian contains them:
\begin{equation}\label{S3orderQRS_A}
H_A^{(3)}(t)=-\sum_{n}\Omega_{n}\Omega_{n+1}\Omega_{n+2}S_{n+2}(t) \int_0^t dt' S_{n+1}(t') \Big(\int_0^{t'}dt'' S_{n}(t'')\Big)|n+3\rangle\langle n| +{\rm H.c.}.
\end{equation}
The contribution of the two-level operators can be easily calculated by noticing that from
\begin{equation}\label{S3order_TL1}
S_{n+2}(t) S_{n+1}(t') S_{n}(t'')=(\sigma_+ e^{i\delta^+_{n+2} t}+\sigma_- e^{i\delta^-_{n+2} t})(\sigma_+ e^{i\delta^+_{n+1} t'}+\sigma_- e^{i\delta^-_{n+1} t'}) (\sigma_+ e^{i\delta^+_{n} t''}+\sigma_- e^{i\delta^-_{n} t''}),
\end{equation}
only the following two terms are not zero (notice that $\sigma_\pm^2=0$)
\begin{equation}\label{S3order_TL2}
S_{n+2}(t) S_{n+1}(t') S_{n}(t'')=\sigma_+ e^{i\delta^+_{n+2} t} e^{i\delta^-_{n+1} t'}e^{i\delta^+_{n} t''} +\sigma_- e^{i\delta^-_{n+2} t} e^{i\delta^+_{n+1} t'}e^{i\delta^-_{n} t''}.
\end{equation}
After calculating the integral we obtain that the Hamiltonian is
\begin{eqnarray}\label{S3orderQRS_A}
H_A^{(3)}(t)=\sum_{n=0}^{\infty}\Omega_{n}\Omega_{n+1}\Omega_{n+2}\Bigg\{\frac{1}{\delta_n^+(\delta_{n+1}^-+\delta_n^+)}\Big(e^{i\delta^{(3)}_{+n}t} - e^{i\delta_{n+2}^+t}\Big) + \frac{1}{\delta^+_n\delta_n^-}\Big(e^{i2(\omega+\gamma)t} - e^{i\delta_{n+2}^+t}\Big)  \Bigg\}\sigma_+|n+3\rangle\langle n| +{\rm H.c} \\ 
+\Omega_{n}\Omega_{n+1}\Omega_{n+2}\Bigg\{\frac{1}{\delta_n^-(\delta_{n+1}^++\delta_n^-)}\Big(e^{i\delta^{(3)}_{-n}t} - e^{i\delta_{n+2}^-t}\Big) + \frac{1}{\delta^-_n\delta_n^+}\Big(e^{i2(\omega-\gamma)t} - e^{i\delta_{n+2}^-t}\Big)  \Bigg\}\sigma_-|n\rangle\langle n+3| +{\rm H.c.}
\end{eqnarray}
where $\delta^{(3)}_{+n}=\delta_{n+2}^++\delta_{n+1}^{-}+\delta_n^+=2\omega+\delta^+_{n+1}$ and $\delta^{(3)}_{-n}=\delta_{n+2}^-+\delta_{n+1}^{+}+\delta_n^-=2\omega+\delta^-_{n+1}$. Ignoring the resonances $\delta^+_{n+2}$ or $\delta^-_{n+2}$ that correspond to the first-order processes, and $2(\omega\pm\gamma)$ which is only zero at the point of the spectral collapse, we are left with 
\begin{equation}\label{S3orderQRS_A2}
H_A^{(3)}(t)=\sum_{n=0}^{\infty}\Omega_{n}\Omega_{n+1}\Omega_{n+2}\Bigg\{\frac{1}{2\delta_n^+(\omega-\gamma)}\sigma_+e^{i\delta^{(3)}_{n+}t}  + \frac{1}{2\delta_n^-(\omega+\gamma)}\sigma_-e^{i\delta^{(3)}_{n-}t}  \Bigg\}|n+3\rangle\langle n| +{\rm H.c.},
\end{equation}
where the three-photon Jaynes-Cummings and anti-Jaynes-Cummings resonances are easily identified as $\delta_{n+}^{(3)}=0$ and $\delta_{n-}^{(3)}=0$ respectively. In a simplified way, Eq.~(\ref{S3orderQRS_A2}) is rewritten as
\begin{equation}\label{S3orderQRS_A3}
H_A^{(3)}(t)=\sum_{n=0}^{\infty}(\Omega^{(3)}_{n+}\sigma_+e^{i\delta^{(3)}_{n+}t}  + \Omega_{n-}^{(3)}\sigma_-e^{i\delta^{(3)}_{n-}t})|n+3\rangle\langle n| +{\rm H.c.},
\end{equation}
where $\Omega^{(3)}_{n+}=g^3\sqrt{(n+3)(n+2)(n+1)}/2\delta^+_n(\omega-\gamma)$ and $\Omega^{(3)}_{n-}=g^3\sqrt{(n+3)(n+2)(n+1)}/2\delta^-_n(\omega+\gamma)$, as shown in the main text.

\section*{Appendix B: Dissipative Effects}

In the main text we have performed the analysis for a closed quantum system, without considering dissipative effects that may appear in a realistic scenario. In the following we show how a dissipative term like the one considered in Eq.~(\ref{Master}) affects the $k$-photon interactions studied in the main text. The dynamics will be described by a master equation of the form
\begin{equation}\label{Master}
\dot{\rho} = -i[H,\rho] + \kappa (2a\rho a^\dagger -a^\dagger a \rho - \rho a^\dagger a),
\end{equation}
where $H$ is the Rabi-Stark Hamiltonian in Eq.~(\ref{QRS1}), $\rho$ is the density matrix describing the state of the system, and $\kappa$ is the coupling constant associated with the dissipative term affecting the bosonic mode. It is expectable that the coherent population exchange will decrease due to the latter. However, in Fig.~\ref{Fig1A} it is shown that the three-photon selective interactions survive for realistic dissipative couplings such as $\kappa/\omega=10^{-3}$ or $\kappa/\omega=10^{-4}$~\cite{Raimond01}. For $\kappa/\omega=10^{-3}$ $(\kappa/g^3=1)$, $P_{e,8}\approx0.1$ is the maximum population obtained for state $|\rm e,8\rangle$, while for $\kappa/\omega=10^{-4}$ $(\kappa/g^3=0.1)$ this value increases to around $0.6$ and more than one oscillation can be observed. In the same manner, to observe a five-photon interaction with $g/\omega=0.1$, a dissipative coupling $\kappa/\omega<10^{-5}$ ($\kappa/g^5<10^{-5}$) will be required.

\begin{figure}
\centering
\includegraphics[width=0.9\linewidth]{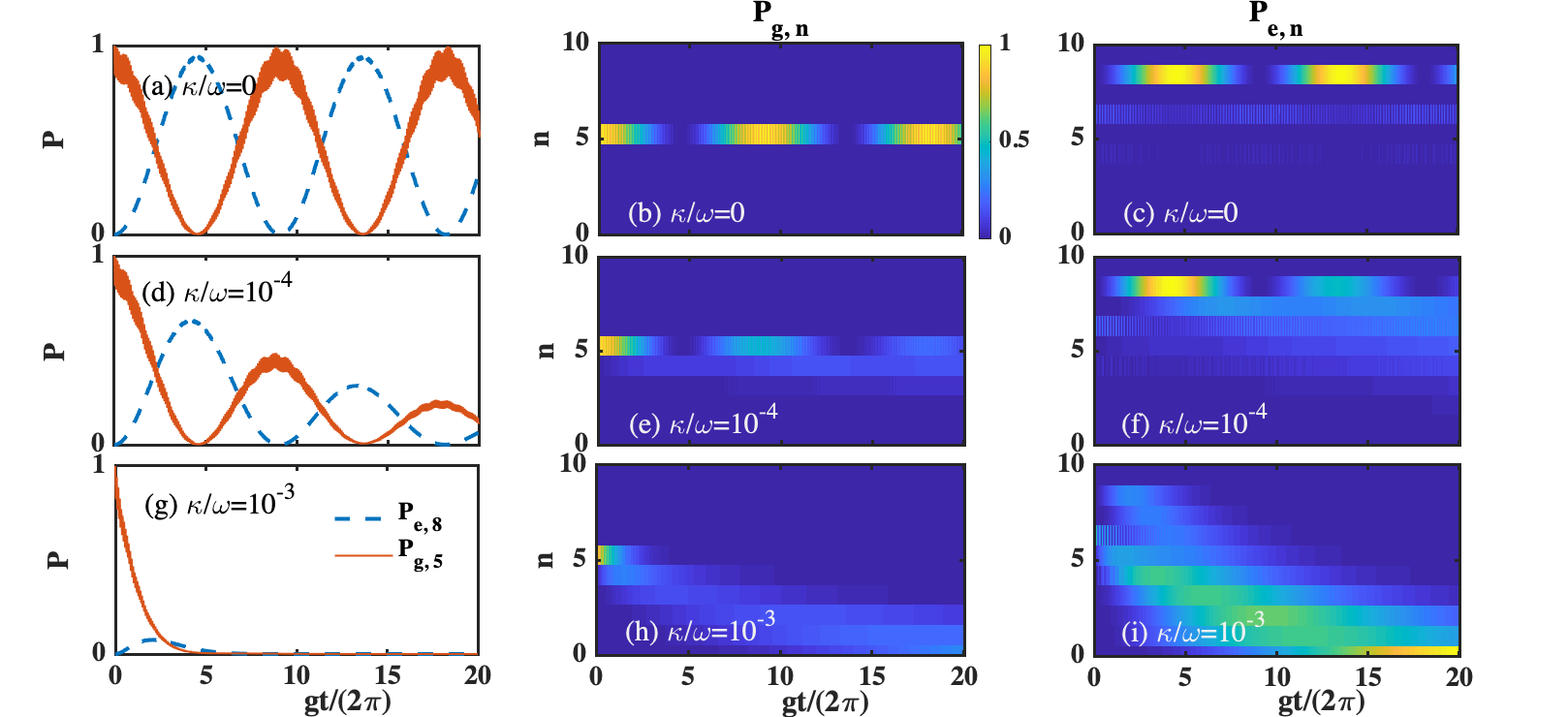}
\caption{Three-photon interactions with dissipative effects for  $g/\omega = 0.1$, $\gamma/\omega = -0.4$, $\omega_0/\omega = 2.317$ and initial state $|\rm g,5\rangle$. In (a), (d) and (g) the time evolution of populations $P_{g,5}$ (solid) and $P_{e,8}$ (dashed) is shown for $\kappa/\omega=0$, $\kappa/\omega=10^{-4}$ and $\kappa/\omega=10^{-3}$ respectively. In (b) and (c), (e) and (f), and (h) and (i) the time evolution of populations $P_{e,n}$ and $P_{g,n}$ is shown, for $\kappa/\omega=0$, $\kappa/\omega=10^{-4}$ and $\kappa/\omega=10^{-3}$ respectively.}  
\label{Fig1A}
\end{figure}

\section*{Appendix C: Derivation of the Rabi-Stark Hamiltonian in trapped ions}
In this section we will explain in detail how to derive the effective Hamiltonian in Eq.~(\ref{TIQRS}) from Eq.~(\ref{Scheme2}),
\begin{equation}\label{SScheme2}
H_{A}(t)=-i\frac{\eta\Omega_r}{2} a \sigma_+ e^{-i\delta_rt}  -i\frac{\eta\Omega_b}{2}  a^\dagger \sigma_+ e^{-i\delta_bt} + e^{i\phi_{\rm S}}g_{\rm S}\sigma_++{\rm H.c.}.
\end{equation}
At this point the vibrational RWA has been applied, however, for a precise understanding of the effective dynamics we need to consider the following additional terms that have been ignored, so that the total Hamiltonian is $H=H_A+H_B$, where
\begin{equation}\label{SSchemeRemaining}
H_{B}(t)=-\frac{\Omega_r}{2} \sigma_+ e^{-i(-\nu +\delta_r)t} -\frac{\Omega_b}{2} \sigma_+ e^{-i(\nu +\delta_b)t} + i\eta e^{i\phi_{\rm S}}\frac{\Omega_{\rm S}}{2}\sigma_+(ae^{-i\nu t}+a^\dagger e^{i\nu t}) +{\rm H.c.}.
\end{equation}
The first two terms are the off-resonant carrier interactions of the red and blue drivings which are usually ignored given that $\Omega_{r,b}\ll \nu$. The last term represents the coupling to the motional mode of the carrier driving. This term does not commute with the first and second terms. In addition, they all rotate at similar frequencies (as $\delta_r,\delta_b\ll\nu$). As a consequence, these terms produce second-order interactions that cannot be ignored as we will see in the following. The second-order effective Hamiltonian is
\begin{equation}\label{SSecondOrder}
H^{(2)}(t)=-i\int_{0}^{t}H(t)H(t')dt'=-i\int_{0}^{t}\Big(H_A(t)+H_B(t)\Big)\Big(H_A(t')+H_B(t')\Big)dt'.
\end{equation}

We are only interested in terms arising from $\int_{0}^{t}H_B(t)H_B(t')dt'$ whose oscillating frequency is $\delta_r$ or $\delta_b$. These are
\begin{eqnarray}\label{SSecondOrderList}
-\frac{\Omega_r}{2}\sigma_+e^{-i(-\nu+\delta_r)t}\int_{0}^tdt'(-i\eta)\frac{\Omega_{\rm S}}{2}e^{-i\phi_{\rm S}}\sigma_- ae^{-i\nu t'} =-\eta\frac{\Omega_{\rm S}\Omega_r}{4\nu}\sigma_+\sigma_-e^{-i\phi_{\rm S}}e^{-i\delta_r t} a, \\
-\frac{\Omega_b}{2}\sigma_+e^{-i(\nu+\delta_b)t}\int_{0}^tdt'(-i\eta)\frac{\Omega_{\rm S}}{2}e^{-i\phi_{\rm S}}\sigma_- a^\dagger e^{i\nu t'} =\eta\frac{\Omega_{\rm S}\Omega_b}{4\nu}\sigma_+\sigma_-e^{-i\phi_{\rm S}}e^{-i\delta_b t} a^\dagger, \\
i\eta\frac{\Omega_{\rm S}}{2}e^{i\phi_{\rm S}}\sigma_+a^\dagger e^{i\nu t}\int_0^t dt' (-\frac{\Omega_r}{2})\sigma_-e^{i(-\nu+\delta_r)t'}=\eta\frac{\Omega_{\rm S}\Omega_r}{4(\nu-\delta_r)}\sigma_+\sigma_-e^{i\phi_{\rm S}}e^{i\delta_r t} a^\dagger, \\
i\eta\frac{\Omega_{\rm S}}{2}e^{i\phi_{\rm S}}\sigma_+a e^{-i\nu t}\int_0^t dt' (-\frac{\Omega_b}{2})\sigma_-e^{i(\nu+\delta_b)t'}=-\eta\frac{\Omega_{\rm S}\Omega_b}{4(\nu+\delta_b)}\sigma_+\sigma_-e^{i\phi_{\rm S}}e^{i\delta_b t} a, \\
-\frac{\Omega_r}{2}\sigma_-e^{i(-\nu+\delta_r)t}\int_{0}^tdt'(i\eta)\frac{\Omega_{\rm S}}{2}e^{i\phi_{\rm S}}\sigma_+ a^\dagger e^{i\nu t'} =-\eta\frac{\Omega_{\rm S}\Omega_r}{4\nu}\sigma_-\sigma_+e^{i\phi_{\rm S}}e^{i\delta_r t} a^\dagger, \\
-\frac{\Omega_b}{2}\sigma_-e^{i(\nu+\delta_b)t}\int_{0}^tdt'(i\eta)\frac{\Omega_{\rm S}}{2}e^{i\phi_{\rm S}}\sigma_+ a e^{-i\nu t'} =\eta\frac{\Omega_{\rm S}\Omega_b}{4\nu}\sigma_-\sigma_+e^{i\phi_{\rm S}}e^{i\delta_b t} a, \\
-i\eta\frac{\Omega_{\rm S}}{2}e^{-i\phi_{\rm S}}\sigma_-a e^{-i\nu t}\int_0^t dt' (-\frac{\Omega_r}{2})\sigma_+e^{-i(-\nu+\delta_r)t'}=\eta\frac{\Omega_{\rm S}\Omega_r}{4(\nu-\delta_r)}\sigma_-\sigma_+e^{-i\phi_{\rm S}}e^{-i\delta_r t} a,  \\
-i\eta\frac{\Omega_{\rm S}}{2}e^{-i\phi_{\rm S}}\sigma_-a^\dagger e^{i\nu t}\int_0^t dt' (-\frac{\Omega_b}{2})\sigma_+e^{-i(\nu+\delta_b)t'}=-\eta\frac{\Omega_{\rm S}\Omega_b}{4(\nu+\delta_b)}\sigma_-\sigma_+e^{-i\phi_{\rm S}}e^{-i\delta_b t} a^\dagger.
\end{eqnarray}
If we assume that $1/(\nu\pm\delta_j)\sim1/\nu $ and reorganize all the terms, we get that the second-order effective Hamiltonian is 
\begin{equation}\label{SSecondOrderEff}
H_B^{(2)}(t)\approx \eta\frac{\Omega_{\rm S}\Omega_r}{4\nu} (ie^{-i\phi_{\rm S}}e^{-i\delta_r t}a+{\rm H.c.})\sigma_z- \eta\frac{\Omega_{\rm S}\Omega_b}{4\nu} (ie^{-i\phi_{\rm S}}e^{-i\delta_b t}a^\dagger+{\rm H.c.})\sigma_z,
\end{equation}
which can be incorporated to the first-order Hamiltonian in Eq.~(\ref{SScheme2}), giving 
\begin{equation}\label{SEffectiveQRS}
H_{\rm eff}(t)=-i(2g^{(1)}_r \sigma_+ -g^{(2)}_{r}e^{-i\phi_{\rm S}}\sigma_z)ae^{-i\delta_rt}  -i(2g_b^{(1)}\sigma_+ +g_{b}^{(2)}e^{-i\phi_{\rm S}}\sigma_z) a^\dagger e^{-i\delta_bt} + \frac{\Omega_0}{2} \sigma_+e^{i\phi_{\rm S}}  -\eta^2\frac{\Omega_{\rm S}}{2}a^\dagger a\sigma_+e^{i\phi_{\rm S}} +{\rm H.c.}, 
\end{equation}
where $g_{r,b}^{(1)}=\eta\Omega_{r,b}/4$ and $g_{r,b}^{(2)}=\eta\Omega_{\rm S}\Omega_{r,b}/4\nu$ and $\Omega_0=\Omega_{\rm S}(1-\eta^2/2)$. Now, if we assume $\phi_{\rm S}=0$ or $\pi$ and move to a frame w.r.t. $\frac{\Omega_{\rm DD}}{2}\sigma_{x}$, we obtain (below $\Omega\equiv\Omega_{\rm DD}$ for clarity)
\begin{eqnarray}\label{SEffectiveQRS2}
H_{\rm eff}^{I}= \frac{\omega_0^{R}}{2}\sigma_+ \mp \eta^2\frac{\Omega_{\rm S}}{2}a^\dagger a\sigma_+ &-&i\Big(g^{(1)}_r (\sigma_x+i\sigma_ye^{- i\Omega t\sigma_x}) \mp g^{(2)}_{r}\sigma_ze^{- i\Omega t\sigma_x}\Big)ae^{-i\delta_rt}  \nonumber\\
 &-&i\Big(g_b^{(1)}(\sigma_x+i\sigma_ye^{- i\Omega t\sigma_x}) \pm g_{b}^{(2)}\sigma_ze^{- i\Omega t\sigma_x}\Big) a^\dagger e^{-i\delta_bt} +{\rm H.c.},
\end{eqnarray}
where $\Omega_{\rm DD}=\pm\Omega_0-\omega_0^R$ for $\phi_{\rm S}=0$ and $\phi_{\rm S}=\pi$ respectively. Using that
\begin{eqnarray}\label{SRotatingPaulis}
\sigma_ye^{- i\Omega t\sigma_x}&=&\cos{(\Omega_{\rm DD}t)} \sigma_y-  \sin{(\Omega_{\rm DD}t)}\sigma_z =\tilde{\sigma}_+e^{ i\Omega_{\rm DD}t} + \tilde{\sigma}_-e^{- i\Omega_{\rm DD}t},\\
\sigma_ze^{- i\Omega t\sigma_x}&=&\cos{(\Omega_{\rm DD}t)} \sigma_z+  \sin{(\Omega_{\rm DD}t)}\sigma_y = -i(\tilde{\sigma}_+e^{ i\Omega_{\rm DD}t} - \tilde{\sigma}_-e^{- i\Omega_{\rm DD}t}),
\end{eqnarray}
where $\tilde{\sigma}_{\pm}=(\sigma_y\pm i\sigma_z)/2$, and that the detunings are chosen to be $\delta_r=\Omega_{\rm DD}+\omega^R$ and $\delta_b=\Omega_{\rm DD}-\omega^R$, Eq.~(\ref{SEffectiveQRS}) is rewritten as
\begin{eqnarray}\label{SEffectiveQRS3}
H_{\rm eff}^{I}= \frac{\omega_0^{R}}{2}\sigma_+ \mp \eta^2\frac{\Omega_{\rm S}}{2}a^\dagger a\sigma_+ + \Big(g^{(1)}_r (-i\sigma_x+\tilde{\sigma}_+e^{i\Omega_{\rm DD}t} + \tilde{\sigma}_-e^{- i\Omega_{\rm DD}t}) \pm g^{(2)}_{r}(\tilde{\sigma}_+e^{i\Omega_{\rm DD}t} - \tilde{\sigma}_-e^{- i\Omega_{\rm DD}t})\Big)ae^{- i\Omega_{\rm DD} t} e^{- i\omega^R t} \nonumber\\
 \Big(g_b^{(1)}(-i\sigma_x+\tilde{\sigma}_+e^{ i\Omega_{\rm DD}t} + \tilde{\sigma}_-e^{- i\Omega_{\rm DD}t}) \mp g_{b}^{(2)}(\tilde{\sigma}_+e^{ i\Omega_{\rm DD}t} - \tilde{\sigma}_-e^{- i\Omega_{\rm DD}t})\Big) a^\dagger e^{- i\Omega_{\rm DD} t}e^{ i\omega^R t} +{\rm H.c.},
\end{eqnarray}
where all terms rotating with $\pm\Omega_{\rm DD}$ or higher can be ignored using the RWA. After the approximation we have that
\begin{equation}\label{SEffectiveQRS4}
H_{\rm eff}^{I}= \frac{\omega_0^{R}}{2}\sigma_x\mp \eta^2\frac{\Omega_{\rm S}}{2}a^\dagger a\sigma_+ + (g^{(1)}_r  \pm g^{(2)}_{r})\tilde{\sigma}_+ a e^{- i\omega^R t}  +(g_b^{(1)} \mp g_{b}^{(2)})\tilde{\sigma}_+  a^\dagger e^{ i\omega^R t} +{\rm H.c.},
\end{equation}
which, in a rotating frame w.r.t $-\omega^Ra^\dagger a$, transforms to
\begin{equation}\label{SEffectiveQRS5}
H_{\rm eff}^{II}= \frac{\omega^R_0}{2}\sigma_x +\omega^R a^\dagger a + g_{\rm JC}(\tilde{\sigma}_+ a+\tilde{\sigma}_- a^\dagger)   +g_{\rm aJC}(\tilde{\sigma}_+  a^\dagger+\tilde{\sigma}_- a) \mp \eta^2\frac{\Omega_{\rm S}}{2}a^\dagger a\sigma_x,
\end{equation} 
where $g_{\rm JC}=\eta\Omega_r(1\pm\Omega_{\rm S}/\nu)/4$ and $g_{\rm aJC}=\eta\Omega_b(1\mp\Omega_{\rm S}/\nu)/4$, depending on the choice of the phase $\phi_{\rm S}=0$ or $\phi_{\rm S}=\pi$.
\end{widetext}

\end{document}